\documentclass[prd,preprint,amsmath,amssymb]{revtex4}%

\usepackage[dvips]{graphics}
\usepackage{array,amsmath,amssymb}
\usepackage{amsmath}
\usepackage{amssymb}

\newcommand{\dis}{\displaystyle}

\begin{document}

%\begin{frontmatter}

%\title{Nonperturbative lattice study of diquark potential}
\title{QCD color interactions between two quarks}

%\author[an]{A.~Nakamura }
%and
%\author[an]{T.~Saito}
%\address[an]{Research Institute for Information Science and Education,
% Hiroshima University, Higashi-Hiroshima 739-8521, Japan}
\author{A.~Nakamura,}
\affiliation{Research Institute for Information Science and Education,
 Hiroshima University, Higashi-Hiroshima 739-8521, Japan}
\author{T.~Saito}
\affiliation{Research Institute for Information Science and Education,
 Hiroshima University, Higashi-Hiroshima 739-8521, Japan}

\begin{abstract}
We study the QCD color interactions between static two heavy quarks
at zero temperature in a quenched $SU(3)$ lattice gauge simulation:
in addition to the standard singlet $q\bar{q}$ potentials,
we calculate octet $q\bar{q}$ potentials, 
symmetric and antisymmetric $qq$ potentials.
It is shown that the antisymmetric $qq$ channel behaves as 
a linearly rising potential at large quark separations.
We further find that the $q\bar{q}$ octet and $qq$ symmetric channels
have the complex dependence on the distance;
at short distances they are repulsive forces,
while at large distances, they show linearly rising feature.
Ratio of string tensions between $q\bar{q}$ singlet and $qq$ 
antisymmetric potentials is described in terms of the Casimir factor.
\end{abstract}

%\begin{keyword}
%% keywords here, in the form: keyword \sep keyword
%% PACS codes here, in the form: \PACS code \sep code
%\sep

%\keywords{
%lattice QCD, color octet channel, color antisymmetric channel, diquark 
%}
%\pacs{12.38.Aw, 12.39.Mk, 12.38.Gc, 11.15.Ha}
%\PACS 12.38.Aw, 12.39.Mk, 12.38.Gc, 11.15.Ha 
%\end{keyword}

%\end{frontmatter}

\maketitle

\section{Introduction}
Nonperturbative study of color QCD forces between static quarks 
is important for understanding quark confinement 
and hadron phenomenology.
The behavior of a heavy-quark potential 
that is defined by the closed gauge-invariant Wilson loop, 
has been studied extensively in lattice simulations \cite{Bali-r};
% the heavy-quark potential is a linearly rising confining potential,  
the heavy-quark potential is a linearly rising potential,  
which can be explained with the Coulomb term and the linear term
with the string tension.
In most previous studies of heavy-quark potentials, a color singlet channel, 
which yields a physical potential, has been investigated.
There are, however, several other color channels between two quarks. 
According to the $SU(3)$ color decomposition for a quark-antiquark sector 
$q\bar{q}$, $ 3 \otimes \bar{3} = 1 \oplus 8 $.
The octet channel is significant for the extensive analysis of 
$J/\psi$ photoproduction \cite{Cacciari,CLEO}.
Furthermore, we have $ 3 \otimes 3 = \bar{3} \oplus 6 $
for a quark-quark sector $qq$,
and in particular, 
the antisymmetric $qq$ channel plays an essential role 
in the phenomenology of penta-quark hadrons \cite{penta-e,penta-t,sasaki},
in which the existence of a highly correlated diquark is assumed
\cite{penta-t}.
Since the usual Wilson loop cannot give
the color-decomposed potentials
separately, the lattice study along this line was considered as
very difficult.

In order to understand quark-gluon plasma (QGP) physics \cite{Gross}, 
%\cite{Gross,Kapusta,Bellac}, 
the finite-temperature behavior of 
the QCD color forces between two static heavy quarks
is studied in lattice simulations.
At finite temperature, one expects that 
% the linearly rising confining color force is weakened
the linearly confining color force is weakened
due to the screening effect in QGP.
Quarks confined in hadrons are expected to move freely in 
the deconfinement phase after the QGP phase transition.
This can be understood in the nonperturbative behavior 
of the screened heavy-quark potentials defined by a Polyakov line 
correlator, the form of which is reduced to a Yukawa-type potential 
with screening masses rather than the linearly rising potential.
We therefore may expect the mass shift 
and/or suppression of heavy quarkoniums 
as an indicator of the QGP phase transition \cite{Miyamura,Matsui}.
However, recent heavy-ion collision experiments
and the lattice QCD simulations \cite{MEM,SGluon}
have indicated that this picture is too simple 
and QGP is a more complex system.
Thus, in recent finite-temperature lattice simulations,
the singlet and octet $q\bar{q}$
potentials and the symmetric and antisymmetric $qq$ potentials 
have been investigated \cite{Nadkarni,CDP1,CDP2,Kacz}.
This is a progress comparing with previous lattice QCD studies,
where the screened heavy-quark potential had focused on 
only a color-averaged channel.
On the other hand, the nonperturbative behavior of 
the $SU(3)$ color-decomposed channels
at zero temperature has not been investigated.

Recently, using $SU(2)$ lattice gauge simulation, 
Greensite, Olejnik and Zwanziger studied the color confinement mechanism 
at zero temperature and 
found that the Coulomb singlet heavy-quark potential 
in Coulomb gauge represents
% a strong linearly rising confining potential \cite{Greensite,Greensite2}, 
a strong linearly rising potential \cite{Greensite,Greensite2}, 
and this was later verified in SU(3) lattice gauge simulation \cite{PPLC3}.
The Coulomb singlet heavy-quark potential is calculated from
partial-length Polyakov lines (PPL) in Coulomb gauge, 
not from the closed gauge-invariant Wilson loop.
PPL is a Polyakov line with a restricted temporal extension
\cite{Greensite,Greensite2,Marinari}. 
Thus, the combinations of the PPL correlators may enable us to carry out 
lattice calculations in the color-decomposed channels at zero temperature.

In this letter, we study the behavior of the color-decomposed $q\bar{q}$
and $qq$ potentials at zero temperature 
in the quenched $SU(3)$ lattice gauge simulation.
Applying the PPL correlator method \cite{Greensite,Greensite2}
to the color-decomposed potentials,
we carry out the $SU(3)$ lattice calculation 
of the octet, symmetric and antisymmetric potentials.
We will present the first nonperturbative study
of the antisymmetric $qq$ potential at zero temperature. 

\section{Partial-length Polyakov line }
%\subsection{Singlet potential with Coulomb gauge}

In this section, we give color decomposed correlators 
between two static heavy quarks and
summarize how to fix the gauge on the lattice.

A partial-length Polyakov line (PPL) can be defined as
\cite{Greensite,Greensite2}
\begin{equation}
L(\vec{x},T) = \displaystyle\prod_{t=1}^{T} U_0(\vec{x},t),
\quad T=1, 2, \cdots, N_t.
\end{equation}
Here $U_0(\vec{x},t)= \exp(iagA_0(\vec{x},t))$
is a $SU(3)$ link variable in the temporal direction and 
$a$, $g$, $A_0(\vec{x},t)$ and $N_t$ represent the lattice cutoff,
the gauge coupling, the time component of gauge potential and
the temporal-lattice size. 
A PPL correlator in the color-singlet channel is given by
\begin{equation}
G_1(R,T) = \frac{1}{3}\left< Tr[L(R,T)L^{\dagger}(0,T)] \right>,\label{pots}
\end{equation}
where $R$ stands for $\arrowvert \vec{x} \arrowvert $.
From Eq. (\ref{pots}) one evaluate a color-singlet potential
on the lattice, 
\begin{equation}
V(R,T) = \log \left[
\frac{G_1(R,T)}{G_1(R,T+a)} \right]\label{pot1}. 
\end{equation}
For the smallest temporal lattice extension, i.e., $T=0$,
we define
\begin{equation}
V(R,0) = - \log [ G_1(R,1) ]\label{pot2}.
\end{equation}

Greensite et al. argued that $V(R,0)$ in Coulomb gauge corresponds 
to a color-Coulomb potential \cite{Greensite,Greensite2}. 
The $V(R,T)$ at $T \rightarrow \infty$ corresponds to a physical potential,
which is usually calculated from 
the Wilson loops at $T \rightarrow \infty$.
These two potentials are expected to satisfy 
Zwanziger's inequality, $V_{phys}(R) \le V_{coul}(R)$ \cite{Zwan}.

Now we extent this PPL correlator method to
the other $SU(3)$ color-decomposed potentials \cite{Nadkarni}.
A color-octet correlator for $q\bar{q}$ is given by 
\begin{equation}
\begin{array}{ccl}
G_8(R,T) &=&\dis \frac{1}{8} \left< TrL(R,T)TrL^{\dagger}(0,T) \right>
          - \dis \frac{1}{24}\left< TrL(R,T)  L^{\dagger}(0,T) \right>,
	 \label{octet}
\end{array}
\end{equation}
and $qq$ correlators in the symmetric and antisymmetric channels are
described as 
\begin{equation}
\begin{array}{ccl}
G_{6}(R,T) & = &\dis \frac{3}{4}\left<TrL(R,T)TrL(0,T)\right>
          + \dis \frac{3}{4}\left<TrL(R,T)L(0,T)  \right>, 
	 \label{symmetric}
\end{array}
\end{equation}
\begin{equation}
\begin{array}{ccl}
G_{\bar{3}}(R,T) &=&\dis \frac{3}{2}\left<TrL(R,T)TrL(0,T)\right>
            -\dis \frac{3}{2}\left<TrL(R,T)L(0,T)  \right>. 
	 \label{antisymmetric}
\end{array}
\end{equation}
In the same way as described in Eqs. (\ref{pot1}) and (\ref{pot2}),  
we will obtain the color-decomposed potentials
in each color channel using the above correlators.
The four potentials between two quarks are classified in terms of 
the Casimir factor on color $SU(3)$ group in the fundamental representation:
$C_{q\bar{q}}^{1}=-\frac{4}{3}$ for a singlet, 
$C_{q\bar{q}}^{8}= \frac{1}{6}$ for an octet, 
$C_{q     q }^{6}= \frac{1}{3}$ for symmetric, 
$C_{q     q }^{\bar{3}}= -\frac{2}{3}$ for antisymmetric; i.e., 
 the $q\bar{q}$ singlet and $qq$ antisymmetric channels yield
 an attractive force, whereas the $q\bar{q}$ octet and
$qq$ symmetric channels give a repulsive force.

Since the color-decomposed potentials defined by the PPL correlators 
do not have gauge-invariant forms, 
we must fix the gauge. We use the Coulomb gauge  
 realized on the lattice as 
\begin{equation}
\mbox{Max} \sum_{\vec{x}} \sum_{i=1}^3
\mbox{ReTr} U_i^{\dagger}(\vec{x},t), 
\end{equation}
by repeating the following gauge rotations:
\begin{equation}
U_i(\vec{x},t) \rightarrow U_i^{\omega}(\vec{x},t)
= \omega^{\dagger}(\vec{x},t)U_i(\vec{x},t) 
\omega(\vec{x}+\hat{i},t),
\end{equation}
where $\omega$ $\in SU(3)$ is a gauge rotation matrix and 
$U_i(\vec{x},t)$ are spatial lattice link variables.
Thus each lattice configuration can be gauge fixed iteratively
\cite{Mandula}.

The Coulomb gauge fixing does not fix a gauge completely,
and one can still perform a time-dependent gauge rotation
 on the Coulomb-gauge fixed links, 
\begin{equation}
\begin{array}{ccl}
U_i(\vec{x},t) &\rightarrow&
\omega^{\dagger}(t) U_i(\vec{x},t) \omega(t), \\
U_0(\vec{x},t) &\rightarrow&
\omega^{\dagger}(t) U_0(\vec{x},t) \omega(t+1) \label{gt}.
\end{array}
\end{equation}
%Thus, some of the PPL correlators, for example,  
Thus, $\mbox{Tr}L\mbox{Tr}L^{\dagger}$ and $\mbox{Tr}LL$
constructed by PPL,
are not invariant under the transformation of Eq. (\ref{gt}). 
Therefore, when performing numerical simulations 
for the octet and two $qq$ correlators including 
$\mbox{Tr}L\mbox{Tr}L^{\dagger}$ and $\mbox{Tr}LL$, 
we additionally implement a global temporal-gauge fixing
on the Coulomb-gauge fixed configuration as
\begin{equation}
\mbox{Max } \frac{1}{V}\sum_{\vec{x},t} \mbox{ReTr} U_0^{\dagger}(\vec{x},t)  
\mbox{ under Eq. (\ref{gt})}, \label{gtg}
\end{equation}
where $V=N_x N_y N_z$ is a spatial lattice volume and 
$N_{x,y,z}$ is the spatial lattice size in each direction. 
Note that this gauge fixing 
does not affect intrinsic Coulomb gauge features.

\section{Simulation results}

We carry out $SU(3)$ lattice gauge simulations
in quench approximation to calculate the color decomposed PPL correlators.
The lattice configurations are generated by the
heat-bath Monte Carlo technique
with a plaquette Wilson gauge action, and
to fix a gauge we adopt the iterative method \cite{Mandula} .

\begin{figure}[htbp]
\begin{center}
\resizebox{9cm}{!}{\includegraphics{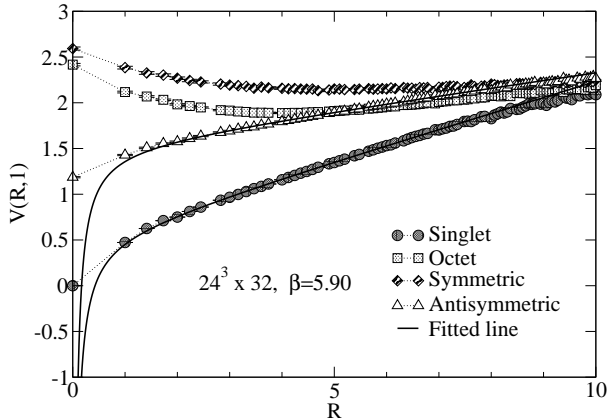}}
%\resizebox{15cm}{!}{\includegraphics{./several.eps}}
\caption{Singlet and octet potentials for $q\bar{q}$ sector and 
symmetric and antisymmetric potentials for $qq$ sector.
The solid curves represent the fitted results for the two attractive channels.
The $\hat{R}$ with physical dimension is set as $\hat{R}=aR$ where
the lattice cutoff $a$ is approximately $0.124fm$ for this lattice simulation. 
%$R=5$ and $R=10$ correspond to approximately $0.62fm$ and $1.24fm$. 
}
\label{4pot}
\end{center}
\end{figure}

Numerical results of the color decomposed potentials with $T=2$ 
are displayed in Fig. \ref{4pot}.
This simulation is done on the $24^3 \times 32$ lattice
at $\beta =5.90$ ($a \sim 0.124 fm$)
and the 200 configurations measured by every 100 steps are
used here. The singlet $q\bar{q}$ and 
the antisymmetric $qq$ potentials 
rise linearly at large distances; i.e., they are
an attractive confining potential.
The octet $q\bar{q}$ and the symmetric $qq$ channels 
show repulsive forces at short distances, while at large distances 
they have the linearly rising attractive force rather than
the large-distance repulsive force.   
This nonperturbative behavior in the adjoint channel is comparable 
with the numerical result in 3d $SU(2)$ gauge theory 
\cite{Philipsen1,Philipsen2}.

In order to analyze the singlet $q\bar{q}$ and the antisymmetric $qq$ 
potentials, we employ the following fitting function, 
\begin{equation}
V(R,T) = C + KR + A/R,\quad A = - \pi/12, \label{fitting}
\end{equation}
where $C$ is a constant and $K$ corresponds to the string tension.
This fitting works well: approximately $\chi^2/ndf \sim O(1)$ for
the fitting range $R$=2-6. 
The results are represented by solid lines in Fig. \ref{4pot}.

\begin{figure}[htbp]
\begin{center}
\resizebox{8cm}{!}{\includegraphics{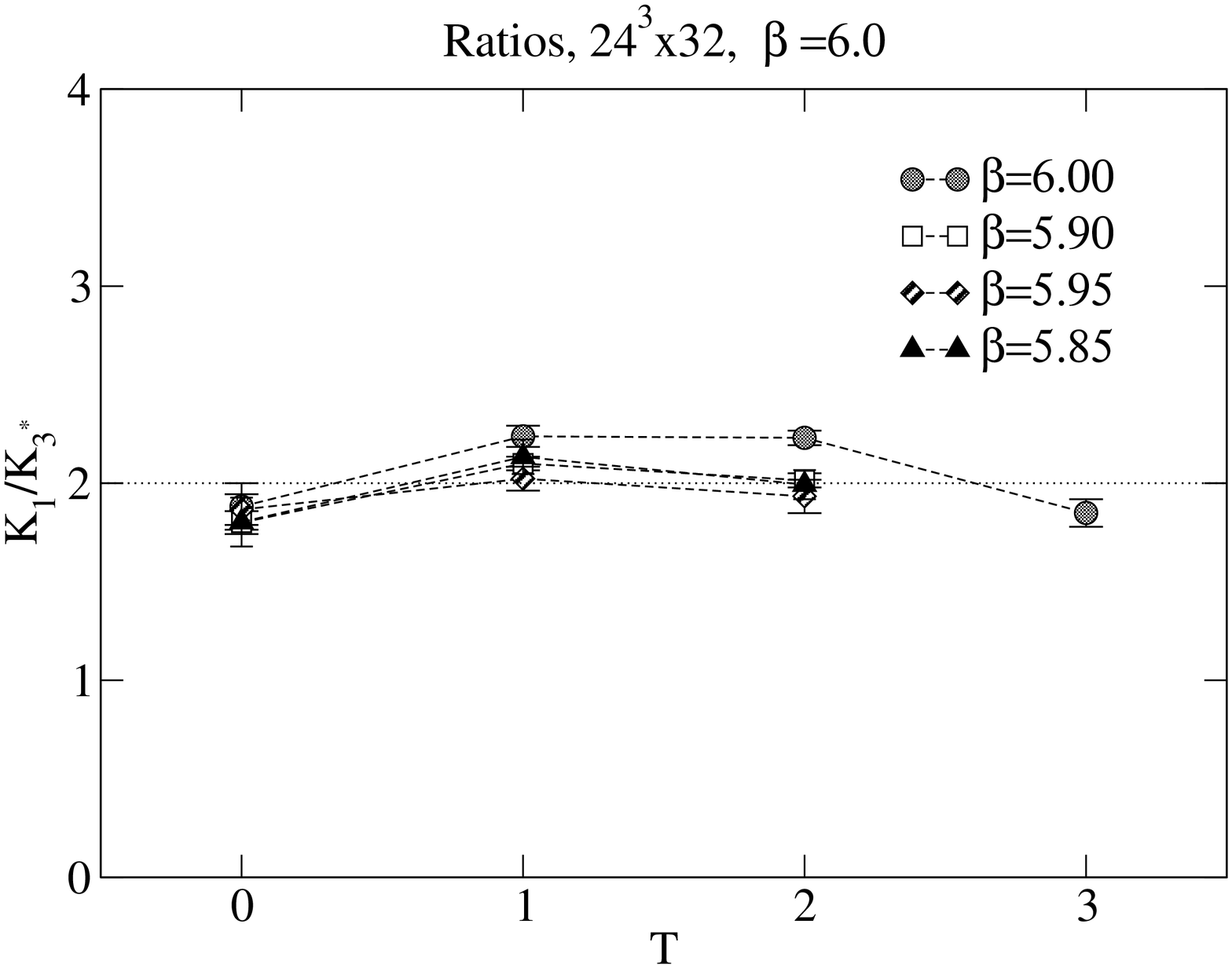}}
%\resizebox{8.0cm}{!}{\includegraphics{./color/ratio.eps}}
\caption{Ratios of string tensions
in the singlet and antisymmetric channels.}
\label{ratio}
\end{center}
\end{figure}

We obtain the string tension $K_{\bar{3}}$ in the antisymmetric channel.
The singlet-string tension $K_1$ approaches 
asymptotically the Wilson-loop string tension \cite{Greensite,PPLC3},
but the $T$ dependence of $K_{\bar{3}}$ is unknown.
Here, we investigate the ratio of the string tensions
obtained from the singlet $q\bar{q}$ and the antisymmetric $qq$ potentials; 
if the string tension itself is universal, then 
the difference between $K_1$ and $K_{\bar{3}}$ can be understood in terms of  
the color $SU(3)$ Casimir factor; that is,
%$K_1/K_{\bar{3}} = (-4/3)/(-2/3) = 2 $.
$K_1/K_{\bar{3}} \rightarrow
C_{q\bar{q}}^{1}/C_{qq}^{\bar{3}}=(-4/3)/(-2/3) = 2 $.
Those ratios calculated at $\beta=5.85-6.00$
are summarized in Fig. \ref{ratio}. 
The dependence of the ratio on $\beta$ and $T$ is weak and they
are consistent with the expected values.
We conclude that
the relation between the singlet $q\bar{q}$ and the antisymmetric
$qq$ potentials
is described in terms of the color $SU(3)$ Casimir 
in nonperturbative regions.

Since the repulsive channels have
the complex dependence in the distance,
we do not carry out a fitting analysis for them.

%\footnote{
%The authors of Refs. \cite{Philipsen1, Philipsen2}
%have calculated singlet and adjoint potentials in 
% 3d SU(2) lattice simulations. However, there is no distinction
%between $q\bar{q}$ and $qq$ potentials in $SU(2)$, because
% both of the potentials have 
%the same irreducible fundamental representation of color $SU(2)$ group.
%}.

%%% koko %%%
%\begin{figure}[htbp]

% \subsubsection{Ratio of string tensions}

%\section{Concluding remarks}
\section{Summary}

We have nonperturbatively studied 
the zero-temperature behavior of the color decomposed potentials
between static two heavy quarks
in the quenched $SU(3)$ lattice gauge simulation 
with the partial-length Polyakov line correlators.

We show that the $qq$ potential in the antisymmetric channel 
rises linearly at large distances and is fitted by the function 
that includes the Coulomb and the string-tension linear terms.
Moreover, the string tensions obtained
from $q\bar{q}$ singlet and $qq$ antisymmetric potentials 
seem to be understood in terms of the Casimir; in other words,
the string tension itself is universal.
As a result, not only the $q\bar{q}$ singlet channel
but also the $qq$ antisymmetric channel
is found to be strongly correlated in nonperturbative regions. 

The octet $q\bar{q}$ and symmetric $qq$ potentials 
show the complex behavior as a function of the distance 
in contrast to the case of the attractive channels:
the $q\bar{q}$ octet and $qq$ symmetric channels
yield a repulsive force at short distances,
while at large distances it seems to be reduced to a linear potential.
The larger lattice simulation therefore may be required to 
deal with the long-range features in the repulsive channels.
%deal with the long-range features in the repulsive channels.
%The contribution of the octet channel becomes important 
%for the detail study of $J/\psi$ photoproduction \cite{Cacciari,CLEO} 
%and the symmetric channel will be needed in an extensive analysis of 
%the hadrons likewise.
%It is also important to investigate the finite volume size effect 
%of color non-singlet channels. 

In the case of color non-singlet channels, it is especially important
to investigate the volume size effect,
because the color non-singlet potentials may include divergent 
parts in the infinite volume limit.

Our numerical simulation has been concentrated on the two quarks system.
With the PPL correlator we may deal with multiquark potentials on the lattice.
The discovery of the penta-quark hadron \cite{penta-e}
promotes the nonperturbative lattice research about 
the multiquark state \cite{sasaki,MQP1,MQP2,Alex}.
The $SU(3)$ lattice simulation of tetra- and penta-quark potentials also 
has been carried out \cite{MQP1,MQP2,Alex}. 
The PPL correlator enables us to construct them without any assumption
to make tentative forms of the multiquark potential, 
although the gauge-fixing procedure is indispensable.

The color decomposed potentials at zero temperature are computed 
in the quenched lattice simulation, and especially 
the behavior of the antisymmetric $qq$ potential should be also
investigated in a dynamical quark simulation.
The PPL correlator ( or the usual Polyakov line correlator )
is applicable to the dynamical-quark lattice simulation. 

\section{Acknowledgment} 

We would like to thank D. Zwanziger for very helpful discussions.
We are also grateful to H. Toki and  V. Dmitra\v sinovi\' c 
for useful comments. 
The simulation was performed on SX-5(NEC) vector-parallel computer 
at the RCNP of Osaka University. 
We appreciate the warm hospitality and support of
 the RCNP administrators.
This work is supported by Grants-in-Aid for Scientific Research from
Monbu-Kagaku-sho (No.11440080, No. 12554008 and No. 13135216).

\end{document}